\documentclass[%
,aps%
 ,secnumarabic%
,amssymb, amsmath,nobibnotes, prd]{revtex4-2}
\usepackage{bm,graphicx}%
\usepackage[colorlinks=true,linkcolor=blue]{hyperref}%
\expandafter\ifx\csname package@font\endcsname\relax\else
 \expandafter\expandafter
 \expandafter\usepackage
 \expandafter\expandafter
 \expandafter{\csname package@font\endcsname}%
\fi
\usepackage{sirimacro}

\begin{document}

\title{Extreme-Value Statistics of the Spin of Primordial Black Holes}%
\author{Siri Chongchitnan}
\email{siri.chongchitnan@warwick.ac.uk}
\affiliation{Warwick Mathematics Institute, University of Warwick, Zeeman Building, Coventry, CV4 7AL, United Kingdom }

\def\simpropto{\lower.2ex\hbox{$\; \buildrel \propto \over \sim \;$}}
\def\ltsim{\lower.5ex\hbox{$\; \buildrel < \over \sim \;$}}
\def\gtsim{\lower.5ex\hbox{$\; \buildrel > \over \sim \;$}}

\author{Joseph Silk}
\email{silk@iap.fr}
\affiliation{Institut d'Astrophysique de Paris, UMR7095:CNRS \& UPMC-Sorbonne University, F-75014, Paris, France}
\affiliation{Department of Physics and Astronomy, The Johns Hopkins University Homewood Campus, Baltimore, MD 21218, USA}
\affiliation{BIPAC, Department of Physics, University of Oxford, Keble Road, Oxford OX1 3RH, United Kingdom}

\date{\today}

\preprint{}
\begin{abstract}
How rare are extreme-spin primordial black holes? We show how, from an underlying distribution of PBH spin, extreme-value statistics can be used to quantify the rarity of spinning PBHs with Kerr parameter close to 1. Using the Peaks-Over-Threshold method, we show how the probability that a PBH forms with spin exceeding a sufficiently high threshold can be calculated using the Generalised Pareto Distribution. This allows us to estimate the average number of PBHs amongst which we can find a single PBH which formed with spin exceeding a high threshold. We found that the primordial spin distribution gives rise to exceedingly rare near-extremal spin PBHs at formation time: for typical parameter values, roughly up to one in a hundred million PBHs would be formed with spin exceeding the Thorne limit. We discuss conditions under which even more extreme-spin PBHs may be produced, including modifying the skewness and kurtosis of the spin distribution via a smooth transformation. We deduce from our calculations that, if indeed asteroid-mass PBHs above the current observational limit on evaporating PBHs of mass $\sim 10^{17}\rm g$ contribute significantly to the dark matter,  it is  likely that some of them  could be near-extremal PBHs.
\end{abstract}
\maketitle

\section{Introduction}

Primordial black holes (PBHs) have long been known as a viable candidate for dark matter and seeds of supermassive black holes (for recent reviews, see \cite{carr, green,villanueva} and references therein). In recent years, PBH mergers have also been put forward as a possible explanation of massive sources of gravitational waves (GWs) observed by the LIGO-VIRGO experiment \cite{abbott, bird, raidal}.

The modelling of PBH mass distribution has been a subject of many previous studies. However, their spin distribution is far less well understood. The spin of a black hole mass $M$ and angular momentum $J$ is characterised by the dimensionless Kerr parameter
\ba a_s={cJ\over GM^2}.\ea The magnitude of $a_s$ is theoretically bounded above by 1 due to the Cosmic Censorship hypothesis \cite{penrose}, since $a_s>1$ yields a naked singularity.  

One of the earliest theoretical predictions of the probability distribution of PBH spins (at formation time) was due to Chiba and Yokoyama \cite{chiba}, who found that PBHs formed during radiation era (due to collapsing overdensities on the Hubble scale) tend to have low spins with $a\lesssim0.4$. Subsequently, de Luca \etal\ \cite{deluca} gave a more sophisticated derivation of a PBH distribution based on peaks theory and also concluded that radiation-era PBH tend to have low spins. The same conclusion was  obtained in further investigations by subsequent authors \cite{mirbabayi, harada}.

On the other hand, forming high-spin PBHs with $a_s$ very close to 1 is not theoretically forbidden. Indeed PBH assembly from particles or fields in the very early Universe can, in principle,  generate high spins \cite{flores}. The existence of a black hole with spin exceeding astrophysical limits (e.g. the Thorne limit $a_s=0.998$ \cite{thorne}) could be construed as  evidence for a primordial origin \cite{arbey, pacheco}.  PBHs in a mass range that otherwise would have evaporated by now, can be partially \cite{arbey} or even fully stabilised \cite{lehmann} against Hawking evaporation by extreme spin or charge. Even if extremely rare at formation, such objects could survive long after their formation epoch and contribute today to observable signals, such as binary mergers in PBH clusters, or delayed stochastic gravitational wave background contributions.   

But exactly \ii{how rare are extreme-spin PBHs?} In this work, we attempt to statistically quantify the rarity of PBHs with $a_s$ close to unity at formation time, and investigate how sensitive this rarity is to changes in the parameters in the underlying PBH formation theory. We will not explicitly consider charge in our discussion of near-extremal black holes, but our statistical treatment of extreme spin values can also be regarded as a proxy for extreme values of charge. In this latter case, the physics can be complicated by pair production, but there are ways to counter such effects, for example by dark photon emission \cite{bai}.

The primary tool of our investigation will be Extreme-Value Statistics (EVS). More precisely, we will employ the Peaks-Over-Threshold (POT) approach which allows us to calculate the probability that a rare PBH forms with spin exceeding a set threshold.  

\section{Review of extreme-value statistics}

\begin{figure}
\begin{center}
\includegraphics[width = 0.9\textwidth]{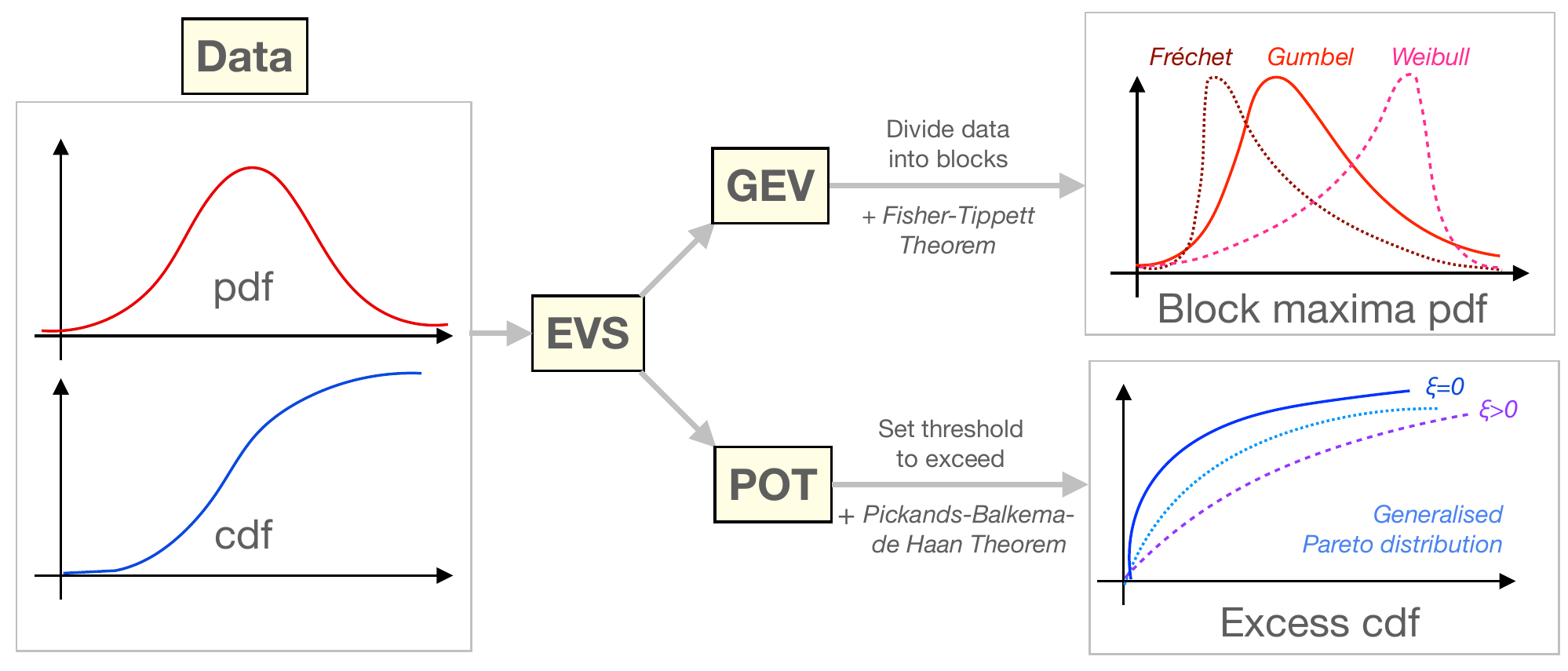}
\caption{Summary of the Extreme-Value Statistics (EVS) pipeline. We start with data described by some underlying probability density function (pdf), or equally, the cumulative density function (cdf). We can then apply one of two EVS methodologies, namely, Generalised Extreme Value (GEV) or Peaks Over Threshold (POT). Both yield different measures of rare events.}
\label{fig_pipeline}
\end{center}
\end{figure}

Extreme-value statistics generally falls into two strands which are summarised below. See Fig. \ref{fig_pipeline} for a graphical summary of the Extreme-Value Statistics pipeline.
 
\bit
\item \ii{Generalised Extreme Value}  (GEV) - also known as the \ii{Block Maxima} approach. The quantity of interest here is the probability distribution of block maxima (or minima), where a block is a set sample size, volume or time period. This involves dividing data into $N$  non-overlapping blocks, and collecting the maximum value from each block (discarding the rest of the data). Under generic assumptions, the large-$N$ limit (after applying a certain scaling) is one of three types: the Gumbel, Fr\'{e}chet or Weibull distribution. This is the result of the Fisher-Tippett-Gnedenko Theorem, which is the key to most applications of EVS (analogous to the Central Limit Theorem).

For examples of previous applications in astrophysics using the block maxima approach, see \cite{davis, waizmann, chongchitnanNG} in the context of massive galaxy clusters, and \cite{kuhnel} in the context of massive PBHs. See \cite{deHaan, gomes} for pedagogical reviews of the GEV approach.

\item \ii{Peaks Over Threshold} (POT) - also known as the \ii{Generalised Pareto} approach. Here the quantity of interest is the probability that an observable exceeds a pre-determined threshold. The analog to the Central Limit Theorem in this approach is the theorem of Pickands \cite{pickands} and Balkema and de Hahn \cite{balkema} which states that for a sufficiently high threshold, the probability of exceeding the threshold can be described by the Generalised Pareto Distribution (more about this in Section \ref{secGPD}).  

As far as we are aware, there have been only a handful of applications of the POT approach to astrophysics, for instance, in \cite{bouillot} in the context of cluster velocities, and in \cite{acero, aschwanden} for solar-physics applications. For more detailed statistical reviews on the POT approach, see \cite{leadbetter, coles, scarrott}.

\eit

The POT approach applied to PBH spin is the main focus of this work.


\section{Statistics of PBH spin}

We now consider the statistics of the spin of PBHs formed during the radiation era. For treatment of the spin of PBHs formed at later times, see \cite{harada,flores}. We will take the pdf of PBH spin to be that given in de Luca \etal  \cite{deluca}. In this formalism, the pdf of PBH spin is characterised by two parameters: $\nu$ and $\gamma$ which are determined by the power spectrum of density perturbations. The parameter $\nu$ is the height of density peaks forming PBHs. Lower values of $\nu$ means the threshold for collapse is reduced, hence leading to an increased abundance of low-mass PBHs. The parameter $\gamma\in[0,1]$ parametrizes the (inverse) width or variance of the power spectrum of density perturbations, where $\gamma=1$ for Dirac-delta power spectrum (giving rise to a monochromatic PBH mass function) whilst smaller $\gamma$ yields a wider range of PBH masses. For example, $\gamma\sim0.82$ for a log-normal power spectrum of density perturbations \cite{deluca}.  Typical values  are $\nu\sim6-9$ and $\gamma\sim0.8-1$. These are the two parameters that we will later vary to determine their effects on the distribution of high-spin PBHs.

The probability density function (pdf) of PBH spins is given as a function of $a_s, \nu, \gamma$ by
\ba P(a_s) = {N_1(a_s, \nu, \gamma) \over N_2(\nu,\gamma)}\ea
where $N_1$ and $N_2$ are rather complicated functions derived from the analysis of tidal torques of overdensity peaks during radiation era. For completeness, we give a compact summary of the analytic form of the pdf in Appendix \ref{appA}. For the rest of this paper, we will refer to the above equation as the \ii{``de Luca's PBH spin pdf"}. We will also work with the corresponding cdf obtained by the usual integration of the pdf.

\section{The Generalised Pareto Distribution for PBH Spin}\lab{secGPD}

We now show that the probability of that a PBH has spin exceeding a high threshold can be approximated using the well-known Generalised Pareto Distribution (GPD). 

Let $X$ be a random variable with the cumulative distribution function (cdf) $F$. In the POT approach, we are interested in the cdf for the \ii{excess distribution} over a pre-determined  threshold $u$, defined by
\ba F_u(x)=P(x\geq X-u\ff|\ff X>u)={F(x+u)-F(u)\over 1-F(u)}.\lab{fu}\ea
The \ii{exceedance}, $x$,  is defined as the upper bound for the difference between the measurements and the threshold $u$ (so if $x=0$ then all measurements never exceed the threshold). We are interested in the rare events where $u$ is set to an atypically high value and the exceedance $x$ is non-negative. In other words, $x$ is defined on the domain $[0, X_F-u]$, where the right endpoint $X_F$ of $F$ is the smallest value such that  $F(X_F)=1$. In other words, the right endpoint is the least upper bound such that the probability of a measurement exceeding $X_F$ is zero.

In our application to PBH spin, $X$ represents the Kerr parameter, $a_s$, and the threshold values $u$ of interest could be one of the following:

\bit
\item $u=0.7$, typical upper range of SMBH spin. 
\item $u=0.8$, typical upper range of the spin of the remnant black hole formed by binary mergers.
\item $u=0.9$, typical upper range of the spin of black holes with mass $\lesssim3\times10^7 M_\odot$.
\item $u=0.998$, the Thorne limit \cite{thorne}, attained by the most extreme astrophysical objects (\eg one possible example being Cygnus X-1  \cite{zhao}).  
\eit
The right endpoint is set to be $X_F=1$ by the Cosmic Censorship hypothesis.

Since the theoretical pdf for the spin is known, in theory the tail of the pdf is also completely known. However, the calculation of probabilities in the tail of the pdf is numerically prohibitive as the extreme precisions needed are hampered by computer round-off errors, made even worse by the large number of operations and integrations involved as evident in Appendix \ref{appA}. We now show that the modelling of the tail using the GPD approach greatly simplifies the problem and can circumvent numerical issues.

In the high-threshold limit, the Pickands-Balkema-de Haan theorem states that if $F$ converges to an extreme-value distribution, then, for sufficiently high threshold values (\ie\ in the limit $u\to x_F$), the tail of the excess cdf can be approximated as
\ba
F_u(x)\simeq G_{\xi,\beta}(x),\ea
where $G_{\xi, \beta}$ is the \ii{Generalised Pareto Distribution} defined by
\ba
G_{\xi,\beta}(x)=\begin{cases}
1-(1+\xi x/\beta)^{-1/\xi}, &\xi\neq0,\\
1-e^{-x/\beta}, &\xi=0.
\end{cases}\lab{gpd}\ea
The parameter $\beta$ is a non-negative function of $u$. $\beta$ is called the \ii{scaling parameter}, and $\xi$ is the \ii{shape parameter}. The GPD is defined where $x\in[0,\infty)$ when $\xi\geq0$, and $x\in[0,-\beta/\xi]$ when $\xi<0$.  Our goal in this section is to numerically calculate $\beta$ and $\xi$ where the underlying pdf is that of the spin of PBHs.

\begin{figure}
\begin{center}
\includegraphics[width = 0.6\textwidth]{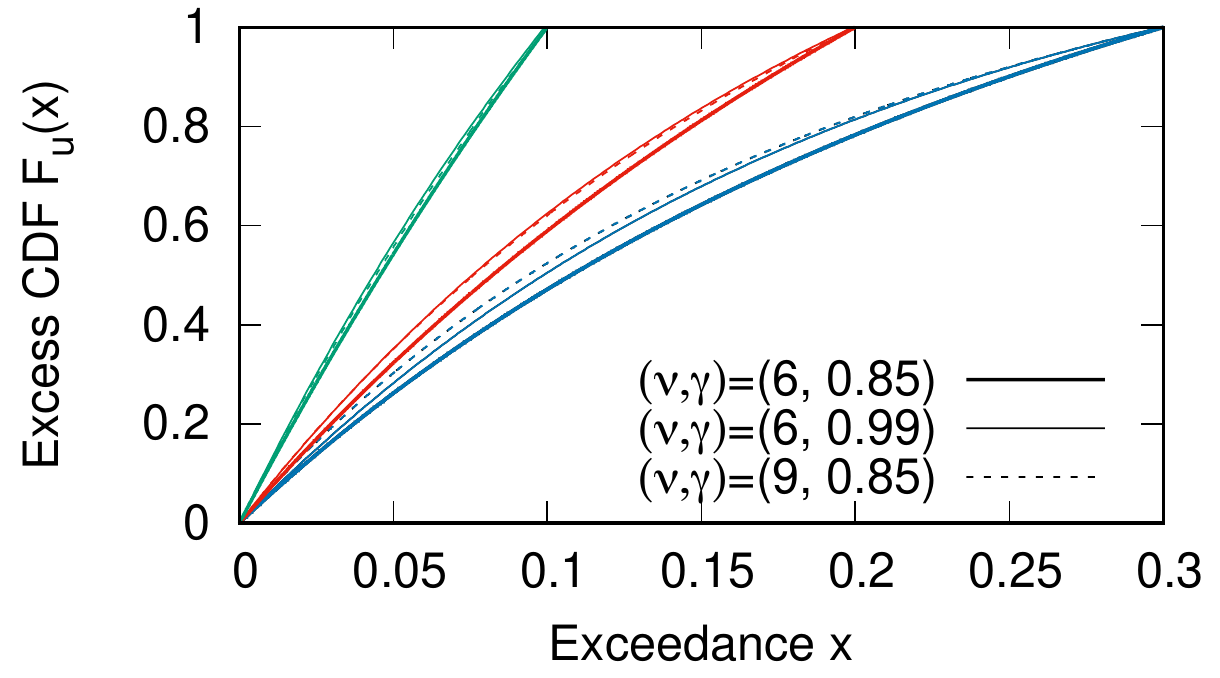}
\caption{The excess cdf $F_u(x)$ defined in Eq. \re{fu}, for $u=0.7$ (blue/rightmost group), 0.8 (red/central group) and 0.9 (green/leftmost group) for various parameter combinations $(\nu,\gamma)$ governing the PBH spin distribution.}
\label{fig_fu}
\end{center}
\end{figure}

Figure \ref{fig_fu} shows the plot of the excess cdf, $F_{u}(x)$, for $u=0.7, 0.8$ and $0.9$ for the following parameter combinations which we will study in this section:
\bee
\item[A.] $\nu=6$, $\gamma=0.85$ 
\item[B.] $\nu=6$, $\gamma=0.99$ 
\item[C.] $\nu=9$, $\gamma=0.85$
\eee

To calculate $\beta$ and $\xi$, we note that the GPD satisfies the relation
\ba \beta+\xi x = {1-G(x) \over G^\pr(x)}.\ea
Therefore, if $F_u(x)$ converges to $G(x)$ for large $u$ and $x$, then $\beta$ and $\xi$ are simply the $y$-intercept and gradient of the  function 
\begin{align}\text{Fit}(x)={1-F_u(x) \over F^\pr_u(x)}.\lab{fit}\end{align}

This observation also gives the following expressions for $\beta$ and $\xi$ in terms of the original pdf and cdf ($f$ and $F$ respectively).
\be
\beta&=\frac{1-F(u)}{f(u)}\\
\xi&=-1-\beta f^\pr(u).
\end{align}

\begin{figure}
\begin{center}
\includegraphics[width = 0.45\textwidth]{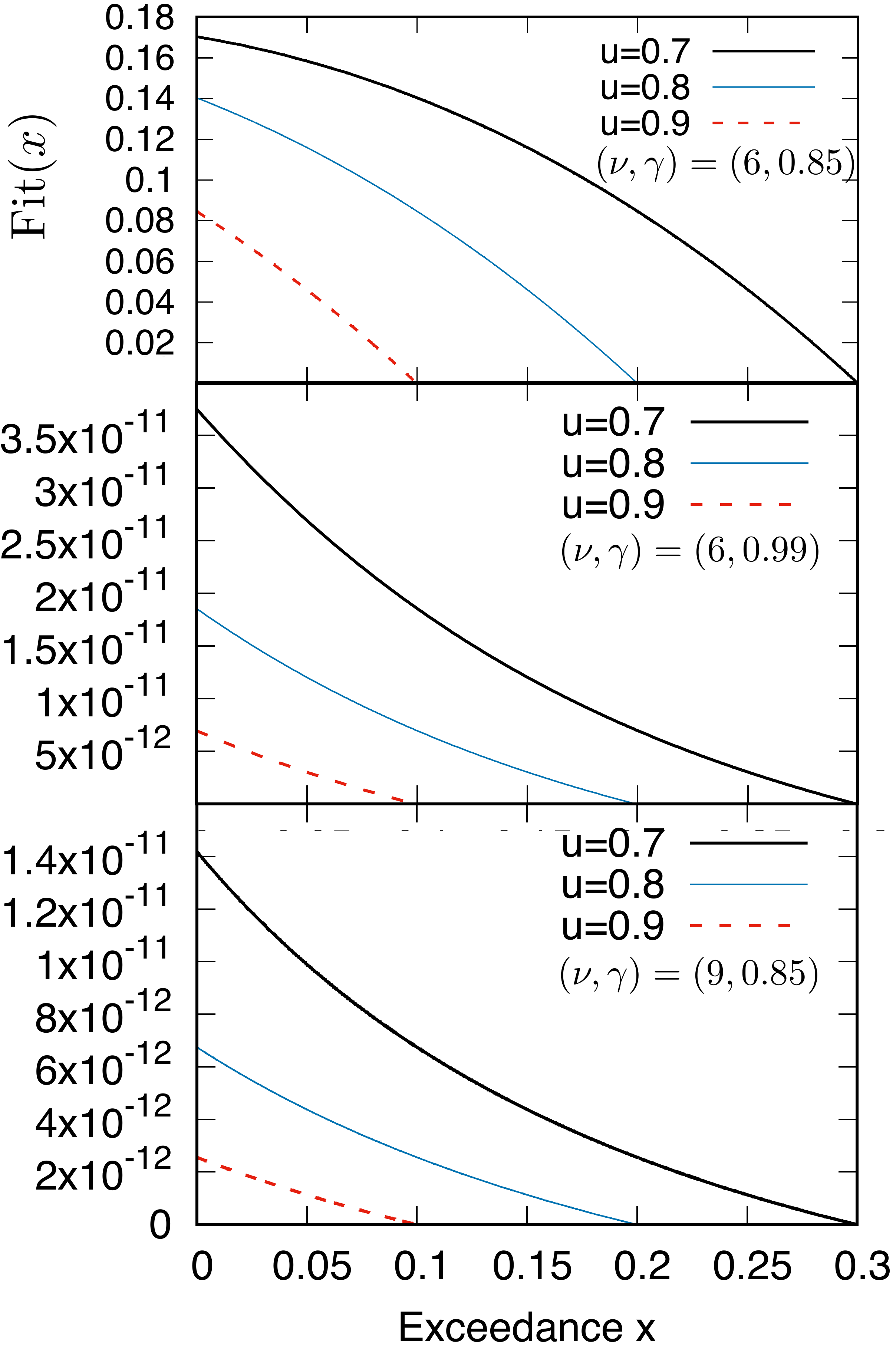}
\caption{Each panel shows the function Fit$(x)$ defined in Eq. \re{fit}, for $u=0.7$ (solid thick line), 0.8 (thin) and 0.9 (dashed) for the  parameter combination $(\nu,\gamma)$ shown. Observe that as $u$ increases in each panel,  Fit$(x)$ becomes more linear, due to the convergence to a generalised Pareto distribution as predicted by the Pickands-Balkema-de Hahn theorem.}
\label{fig_fit}
\end{center}
\end{figure}

Figure \ref{fig_fit} shows the graphs of Fit$(x)$ for various parameter combinations. We observe that as $u$ increases in each panel,  Fit$(x)$ indeed becomes more linear. If $F_u(x)\to G(x)$ then it can be shown that $\beta(u)$ is a linear function of $u$, and $\xi$ is a constant independent of $u$. In our case, the equation of the dashed lines is Fit$(x)=\beta+\xi x.$ We now numerically investigate the approximations of $\beta$ and $\xi$, which together completely characterise the GPD.

\subsection*{Lowest-order approximation}

At lowest-order approximation, \ie\ in the very high-threshold limit ($u\gtrsim0.9$), we find
\be
\beta(u)&\approx 1-u\lab{limb}\\
\xi&\approx-1.\lab{limb2}
\end{align}
regardless of the values of $\nu$ and $\gamma$. Substituting these into \re{gpd}, we find that, for $u\simeq 1$, 
\ba F_u(x)\simeq {x\over 1-u}.\ea Correspondingly, this means that the cdf $F(x)$ is linear in the tail, and that the spin pdf is roughly flat. For our chosen parameter combinations, the above approximations are accurate to $1\%$ for threshold values $u\in(0.95, 1)$ and therefore is sufficient for studying the occurrences of PBHs violating the Thorne limit ($a_s=0.998$). We note that attempting to evaluate the de Luca's spin cdf for $a_s\gtrsim0.998$ without the POT approach is problematic since floating-point errors conspire to give unity, giving us no information about the tail.   

\subsection*{Next-order approximation}

For lower thresholds ($0.7\lesssim u\lesssim0.9$), let us obtain the next-order approximation of $F_u(x)$. The algebraic requirement $\beta(1)=0$ and the large-$u$ limit \re{limb} imply that the next-order approximation of $\beta(u)$ is quadratic. We can express $\beta$ as
\ba\beta(u)=(1-u)(1-C(1- u)).\lab{next}\ea
This parametrizes the $y$-intercept of the line $y=$Fit$(x)=\beta+\xi x$. Together with the fact that Fit$(1-u)=0$, we have the following expression for $\xi$
\ba \xi=-1+C(1-u),\lab{next2}\ea
valid for small $(1-u)$. Therefore there is only one parameter $C$ to determine from least-square fitting, which then determines both $\beta$ and $\xi$. Note that the limits \re{limb}-\re{limb2} are recovered from these next-order expressions.

Table \ref{tab2} shows the parameter values for $C$ for various combinations of $(\nu,\gamma)$ (including extended values of $\nu$ down to 2). We found the next-order approximations to be accurate to $1\%$ for threshold values $u\in(0.8, 1)$. In particular, For $(\nu,\gamma)=(6,0.85)$ the formula remains applicable at subpercent accuracy for $u\geq0.7$.

\begin{table}[h]
\caption{The parameter $C$ for the next-order approximations of GPD parameters $\beta$ and $\xi$  defined in Equations \re{next}-\re{next2}, for various combinations of PBH-spin distribution parameters, $(\nu,\gamma)$.}
\begin{center}
\begin{tabular}{|c|c|c|}
\hline
Model &\hskip 0.3cm $(\nu, \gamma)$ \hskip 0.2cm\, & \hskip 0.3cm $C$ \hskip 0.2cm\,\\
\hline
A&(6, 0.85) & $1.53$\\
B&(6, 0.99) & $2.07$\\
C&(9, 0.85) & $1.97$\\
\hline
\end{tabular}\quad
\begin{tabular}{|c|c|}
\hline
\hskip 0.3cm $(\nu, \gamma)$ \hskip 0.2cm\, & \hskip 0.3cm $C$ \hskip 0.2cm\,\\
\hline
(5, 0.85) & $1.38$\\
(4, 0.85) & $1.27$\\
(3, 0.85) & $1.18$\\
(2, 0.85) & $1.11$\\
\hline
\end{tabular}
\end{center}
\label{tab2}
\end{table}%

\begin{figure}
\begin{center}
\includegraphics[width = 0.45\textwidth]{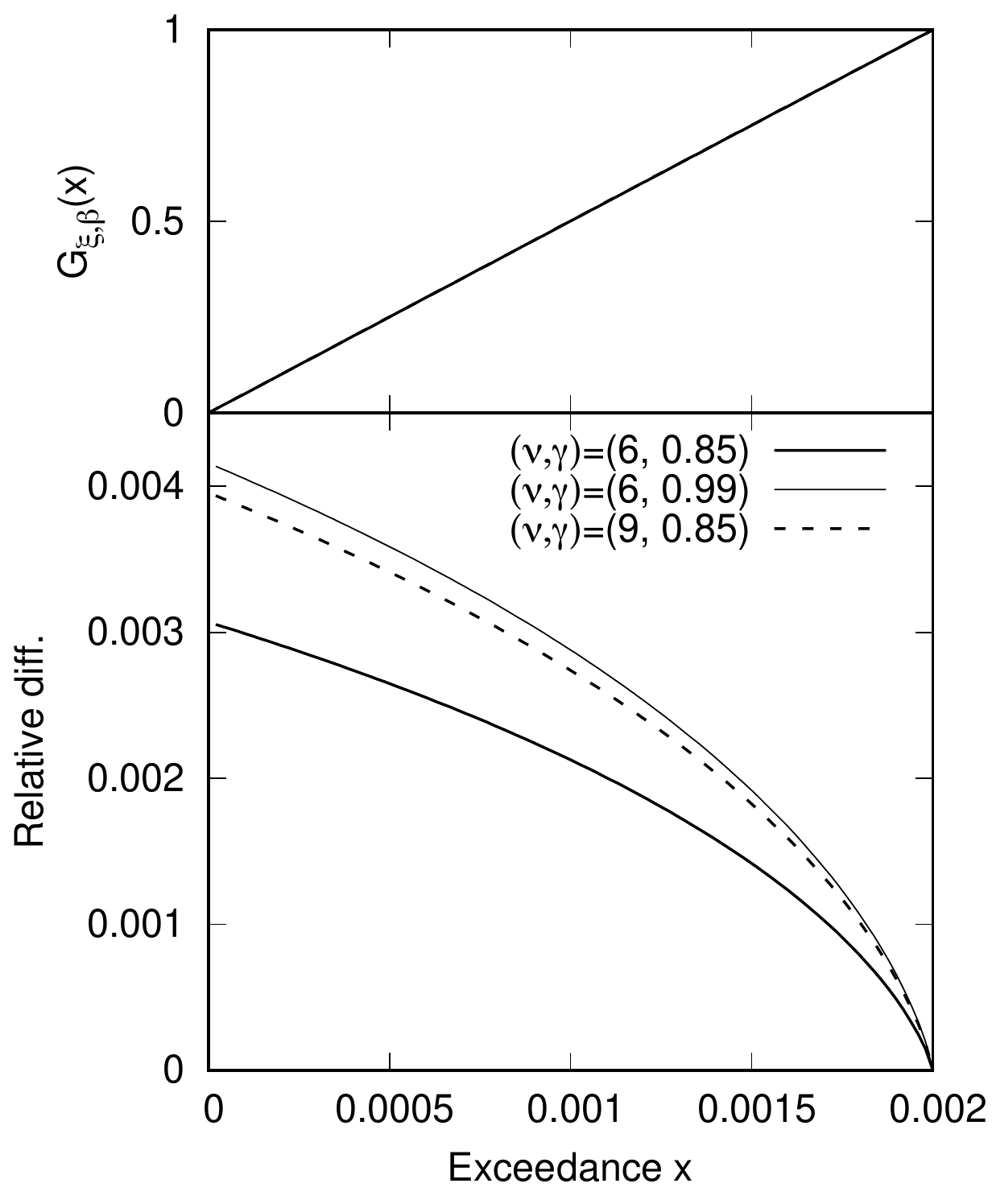}
\caption{The Generalised Pareto model of the excess cdf of PBH spins exceeding the Thorne limit $u=0.998$. All 3 curves curves are visually indistinguishable on the top panel, which shows $G_{\xi,\beta}(x)$. The lower panel shows the relative difference between the next-order and lowest-order approximations of the GPD (Eq. \ref{rel}).}
\label{fig_double}
\end{center}
\end{figure}

Finally, we test the accuracy of the lowest-order versus the next-order approximations for the Thorne limit case. Fig \ref{fig_double} shows the excess cdf, $F_u(x)$ for the Thorne limit $u=0.998$, using the next-order approximation. We note again that for such a high threshold, the de Luca's spin cdf for the case $\nu=9$ cannot be evaluated accurately due to floating-point errors, but this issue is circumvented by the GPD modelling. The bottom panel shows the relative difference between the next-order and lowest-order approximations, \ie\
\begin{align} \text{Relative difference}={G_{\xi,\beta} (\text{next order}) \over G_{\xi,\beta} (\text{lowest order})}-1.\lab{rel}\end{align}
We see that the relative improvement in going from lowest to next order approximation given such a high threshold is well below subpercent. We conclude that in the case of the Thorne limit,  the lowest-order approximation does indeed provide an accurate model of the GPD tail.

\section{Return levels}

Extreme-Value Statistics can help us quantify the rarity of an event via the concept of the \ii{return level}, which, in the simplest term, is embodied in statements such as ``\ii{the event occurs once in $N$ years}", or ``\ii{the event is seen once every $N$ measurements}", depending on which EVS methodology is used:
\bit
\item In the GEV approach, the return level $z_N$ is the value that is expected to be exceeded once every $N$ years (or $N$ blocks). For example, we could  estimate the maximum energy of solar flares expected in a  100-year period, or ask how often do the most powerful X-class solar flares occur \cite{tsiftsi}.
\item In the POT approach, the return level $x_N$ is the value that is exceeded on average once every $N$ measurements.
\eit
Let us discuss how $x_N$ can be calculated in the POT approach. The probability of exceedance of a random variable $X$ given that it exceeds a high threshold $u$ can be written in terms of GPD parameters as
\begin{align} 
\text{P}(X>x\,|\,X>u)&= 1- \text{P}(x\geq X\,|\,X>u)\notag\\
&=  1- \text{P}(x-u\geq X-u\,|\,X>u)\notag\\
&=  1- F_u(x-u)\qquad\quad\text{(using Eq. }\ref{fu})\notag \\
&\approx  1- G_{\xi,\beta}(x-u)\qquad\text{ (assuming the P-B-dH theorem)}\notag\\
&=\bkts{1+\xi\bkt{x-u\over \beta}}^{-1/\xi},\end{align}
where $x>u$ and we have assumed $\xi\neq0$ (substantiated by the results in the previous section).

Let $\zeta_u=\text{P}(X>u)=1-F(u)$, which is the probability of exceeding the threshold. Bayes' Theorem states that
\[\text{P}(X>x)\cdot \text{P}(X>u | X>x)=\text{P}(X>u)\cdot\text{P}(X>x | X>u) \]
The second probability is 1 since $x>u$. Setting $P(X>x)={1\over N}$ and substituting our results so far gives an equation satisfied by the return level $x_N$, \ie\ the value that is exceeded, on average, once every $N$ observations. 
\begin{align} \zeta_u\bkts{1+\xi\bkt{x_N-u\over \beta}}^{-1/\xi}={1\over N}.\end{align}
Solving for the return level:
\begin{align} x_N=u+{\beta\over\xi}\bkts{(N\zeta_u)^\xi-1}.\end{align}



\begin{figure}
\begin{center}
\includegraphics[width = 0.45\textwidth]{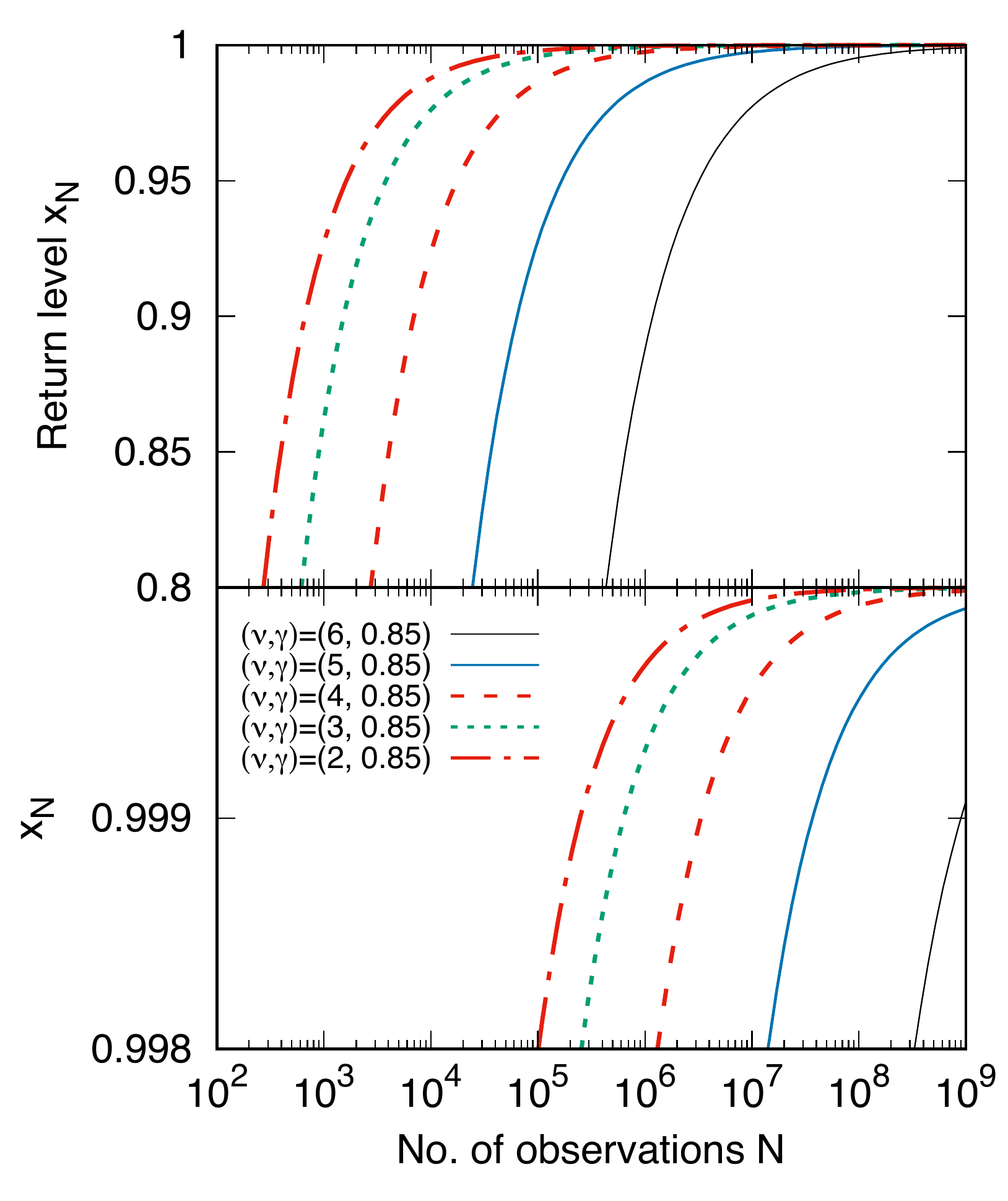}
\caption{The return-level plot, showing the sample size $N$ in which we would find a single PBH with spin exceeding $x_N$.  We assume the parameters $\gamma=0.85$ and $\nu=2-6$. In the upper panel, $x_N\geq0.8$, whilst the lower panel zooms into the region where  $x_N$ exceeds the Thorne limit $0.998$.}
\label{figxn}
\end{center}
\end{figure}

In Fig. \ref{figxn}, the upper panel shows the return level $x_N$ plotted against $N$ for the threshold $u=0.8$, using the next-order approximation \re{next}-\re{next2} for the GPD.  Using the PBH-spin distribution of de Luca \etal\ with standard parameters ($\nu=6-9$), we find that only one PBH amongst a huge number of PBHs would be formed with extremal spin. For instance, for the threshold  $a_s=0.8$, one such PBH would form for every  $10^5-10^6$ PBHs (assuming models A and B), and $10^{11}$ (model C). These numbers become more extreme for a single Thorne-limit violation ($u=0.998$, lower panel in Fig. \ref{figxn}): reading off from the graph, we see that $N\sim10^8$ for models A and B, and $\sim10^{14}$ for model C (not shown). These numbers quantify how exceedingly rare it is to form extreme-spin PBHs. These are the central results of this paper.

We also plotted the return levels for models with an extended range of $\nu$ ($2-5$). For these models, we see expect $N$ of order 100 for a return level $x_N\sim0.8$, and $10^5$ for the Thorne limit. Recall that $\nu$ controls the height of overdensity peaks that end up in PBHs, decreasing $\nu$ would naturally overpredict the abundance of PBHs, but in any case the graphs give us an indication that decreasing $\nu$ further is a limited way to produce extreme-spin PBHs.

\section{Modifying the skewness and kurtosis of Spin pdf}

The results in the previous section show that the de Luca's PBH spin distribution generally gives rise to exceedingly rare extreme-spin PBHs. We ask: what modifications could one perform on the underlying PDF that might increase the chances of observing extreme-spin PBHs? It is likely that interactions between PBHs could change the primordial PBH distribution. Merger processes could, for instance, skew the spin distribution towards higher values of $a_s$. In this section, we systematically modify the skewness and kurtosis of spin distribution to mimic post-formation effects. Adding kurtosis and skewness are generic extensions of PBH formation modelling. We show that their inclusion allows a considerable boost in the return levels of extreme-spin PBHs

Let $X$ be a random variable with pdf $f$, mean $\mu=\bkta{X}$ and variance $\sigma^2=\bkta{X^2}-\bkta{X}^2$, the skewness and kurtosis are defined by
\ba
\text{Skewness}&={\bkta{(X-\mu)^3}\over \sigma^3}={\bkta{X^3}-3\mu\bkta{X^2}+2\mu^3\over \sigma^3},\\
\text{Kurtosis}&={\bkta{(X-\mu)^4}\over \sigma^4}\nn\\
&={\bkta{X^4}-4\mu\bkta{X^3}+6\mu^2\bkta{X^2}-3\mu^4\over \sigma^4}.
\ea

For example, the normal distribution has skewness 0 and kurtosis 3. Table \ref{t22} shows the values of the skewness and kurtosis for some of the models considered in \S\ref{secGPD}.

\begin{table}[h!]
\caption{The skewness and kurtosis for some models of the PBH spin distribution (to 3 significant figures).}
\begin{center}
\begin{tabular}{|c|c|c|}
\hline
$(\nu,\gamma)$ & Skewness & Kurtosis\\
\hline
Model A: $(6, 0.85)$  &  8.06 & 251 \\
Model B: $(6, 0.99)$  & $2.45$ & $76.3$\\
Model C: $(9, 0.85)$ & $2.27$ & $14.1$\\
$(2, 0.85)$ & $4.12$ & $24.3$\\
\hline
\end{tabular}
\end{center}
\lab{t22}
\end{table}

We now consider a specific method  for generalizing  our calculations to incorporate control on the skewness and kurtosis. Given the spin cdf, $F$, parametrized by $(\nu, \gamma)$, we introduce the \ii{sinh-arcsinh (SAS) transformation}  to introduce additional skewness and kurtosis to the pdf. This procedure was first introduced by Jones and Pewsey \cite{jones} in the attempt to smoothly introduce asymmetry to the normal distribution. This was done via a 2-parameter transformation of the cdf, $F\to F_{\eps,\delta}$, where
\ba
F_{\eps,\delta}(\hat{x})&=F(\sinh\bkt{\delta\sinh^{-1}\hat{x}-\eps}),\lab{saseq}\\
\hat{x}&={x-\mu\over \sigma}.
\ea
The parameters $\eps$ and $\delta$ affect skewness and kurtosis respectively (if the pdf is symmetric). Note that when $(\eps,\delta)=(0,1)$, we obtain the identity transformation. The perturbed pdf can be obtained by differentiating the perturbed cdf,  yielding
\ba f_{\eps,\delta}(\hat{x})=f\bkt{\sinh\bkt{\delta\sinh^{-1}\hat{x}-\eps}}\cosh\bkt{\delta\sinh^{-1}\hat{x}-\eps}{\delta\over\sqrt{1+\hat{x}^2}}.\ea

\begin{figure}
\begin{center}
\includegraphics[width = 0.6\textwidth]{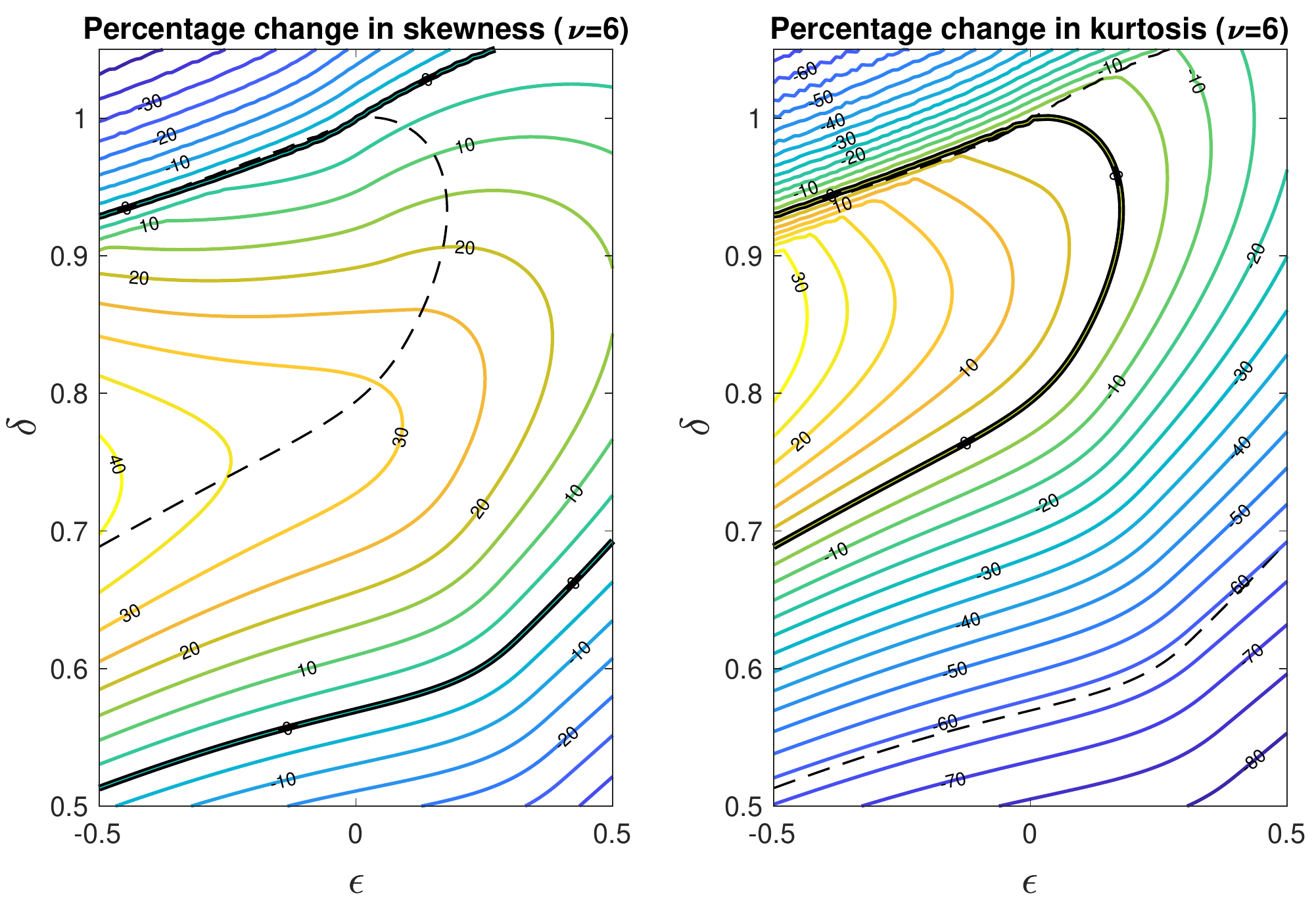}
\includegraphics[width = 0.6\textwidth]{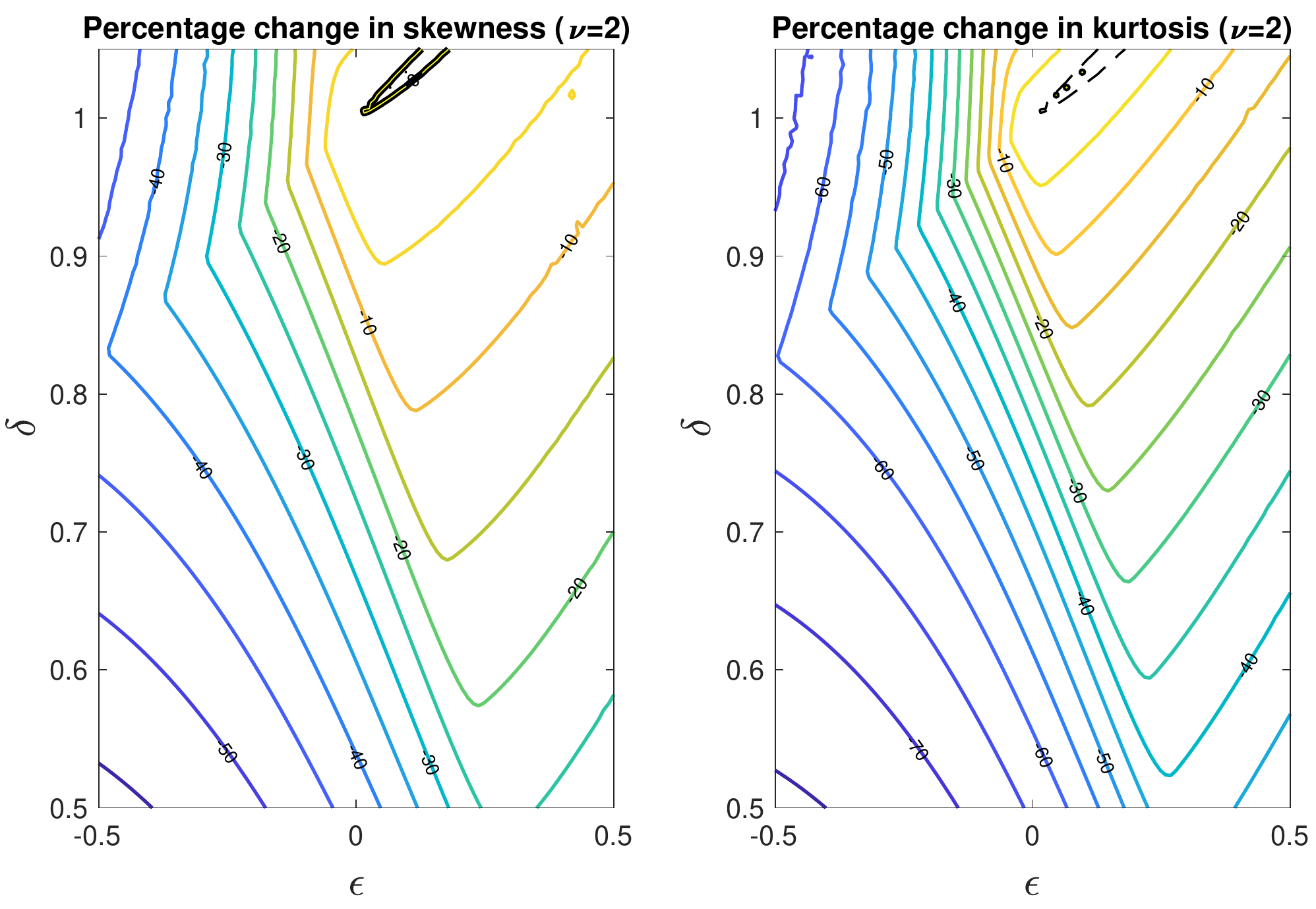}
\caption{Percentage change in skewness (left) and in the kurtosis (right) in the PBH probability density distribution model with $(\nu,\gamma)=(6, 0.85)$ (top) and $(\nu,\gamma)=(2, 0.85)$ (bottom), using the \ii{sinh-arcsinh} transformation parameters $(\eps, \delta).$ In the skewness figure, the thick black line shows the combinations of $(\eps,\delta)$ yielding no change in skewness, whilst the dashed line shows those yielding no change in the kurtosis (vice versa for the kurtosis figure).}
\label{figcontour}
\end{center}
\end{figure}

We demonstrate how this technique can be applied to the spin pdf as follows. First, we scan the parameter space of $\eps$ and $\delta$ to see how changing each variable affects the skewness and kurtosis (since the spin pdf is asymmetric, $(\eps,\delta)$ do not necessarily alter the skewness and kurtosis separately). Fig. \ref{figcontour} shows how the contour plots of $(\eps,\delta)$  and the corresponding percentage changes in the skewness (left panels) and kurtosis (right panels) relative to the identity transformation. We plotted the contours for a typical value $\nu=6$ (Model A, top panels), and an atypically low value $\nu=2$ (bottom panels) for comparison. 

From the contour plots, we were able to identify the combinations of $(\eps,\delta)$ which would affect only the skewness or kurtosis separately (thick black contour lines).

We experimented with various combinations of $(\eps,\delta)$ along the curves of constant skewness and constant kurtosis, but found that changing them individually by $\sim\pm20\%$ resulted in negligible effects on the return levels.  Instead, we found that an appreciable amount of boost to the return level was obtained when both skewness and kurtosis are varied together.

We shall demonstrate this using two transformations: $(\eps,\delta)=(0.25, 0.75)$ and $(\eps,\delta)=(0.5, 0.5)$. The effects on the skewness and kurtosis are summarised in Table \ref{tab} below.  

\begin{table}[h!]
\caption{The parameters $(\eps,\delta)$ of two SAS transformations acting on the spin pdf, and their effects on the skewness and kurtosis compared with the identity transformation $(\eps,\delta)=(0,1)$.}
\begin{center}
\begin{tabular}{|c|c|c|}
\hline
$(\nu,\gamma)=(6, 0.85)$ &$\Delta$ Skewness &$\Delta$ Kurtosis\\
\hline
$(\eps,\delta)$=(0,1)&  0\% & 0\% \\
$(\eps,\delta)$=(0.25,0.75)  & $+22.8\%$ & $-28.6\%$\\
$(\eps,\delta)$=(0.5,0.5) & $-33.4\%$ & $-84.7\%$\\
\hline
\end{tabular}\qquad
\begin{tabular}{|c|c|c|}
\hline
$(\nu,\gamma)=(2, 0.85)$ &$\Delta$ Skewness &$\Delta$ Kurtosis\\
\hline
$(\eps,\delta)$=(0,1)&  0\% & 0\% \\
$(\eps,\delta)$=(0.25,0.75)  &$-13.2\%$ & $-26.4\%$\\

$(\eps,\delta)$=(0.5,0.5) & $-28.9\%$ & $-49.0\%$\\
\hline
\end{tabular}
\end{center}
\label{tab}
\end{table}%

\begin{figure}
\begin{center}
\includegraphics[width = 3.4in]{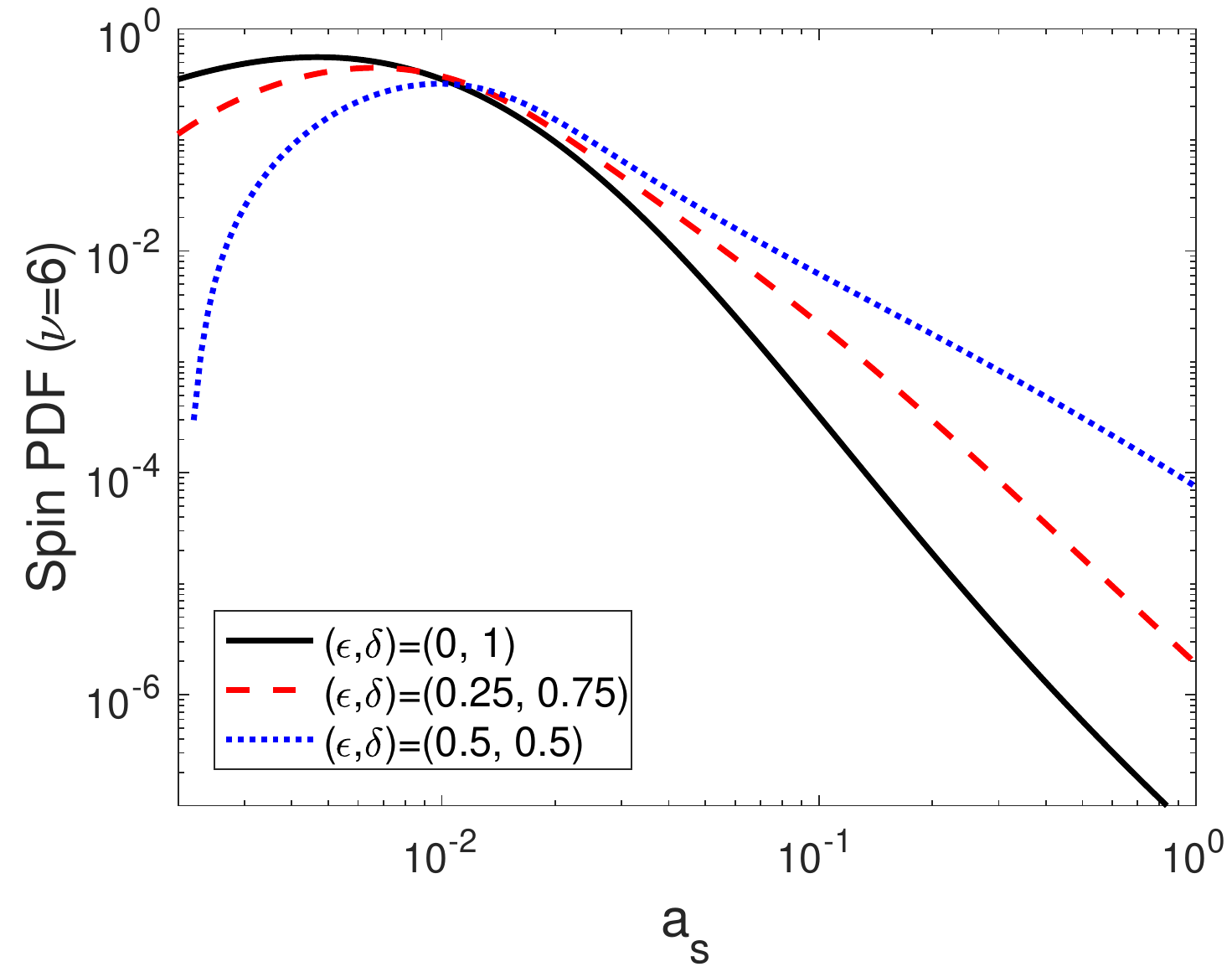}
\includegraphics[width = 3.4in]{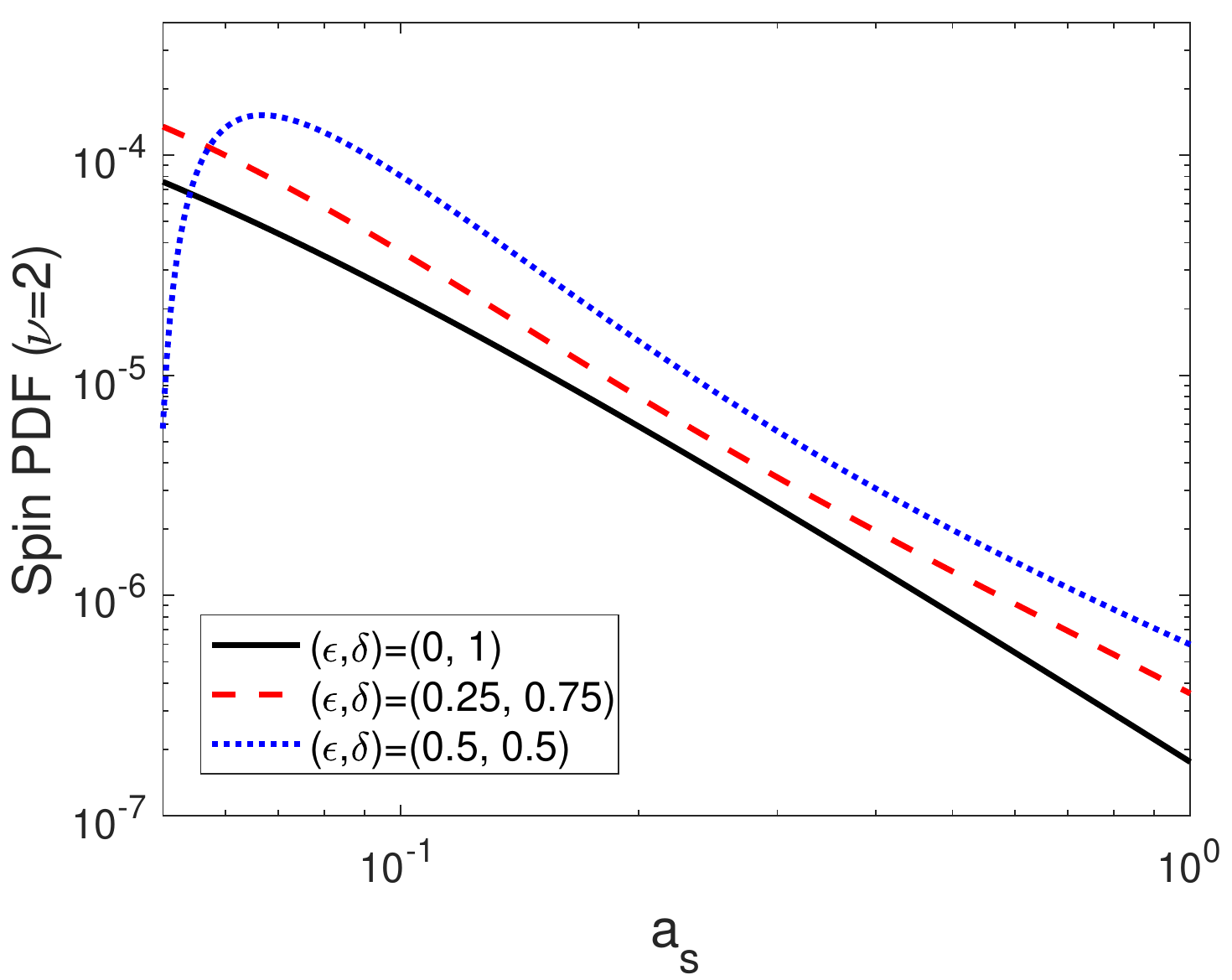}
\caption{The solid/black lines show the PBH spin pdf with $(\nu,\gamma)=(6, 0.85)$ (left) and  $(\nu,\gamma)=(2, 0.85)$ (right), together with their images under two SAS transformations: $(\eps,\delta)=(0.25, 0.75)$ (dashed/red) and $(\eps,\delta)=(0.5, 0.5)$ (dotted/blue). The probability of forming PBHs with extreme spins increases with these transformations.}
\label{figSASpdf}
\end{center}
\end{figure}

The effects of these transformation on the spin pdf are shown in Fig. \ref{figSASpdf}. We see that these transformations suppress the low-spin probabilities in favour of high-spin values around $a_s\sim1$.

\begin{figure}
\begin{center}
\includegraphics[width = 0.9\textwidth]{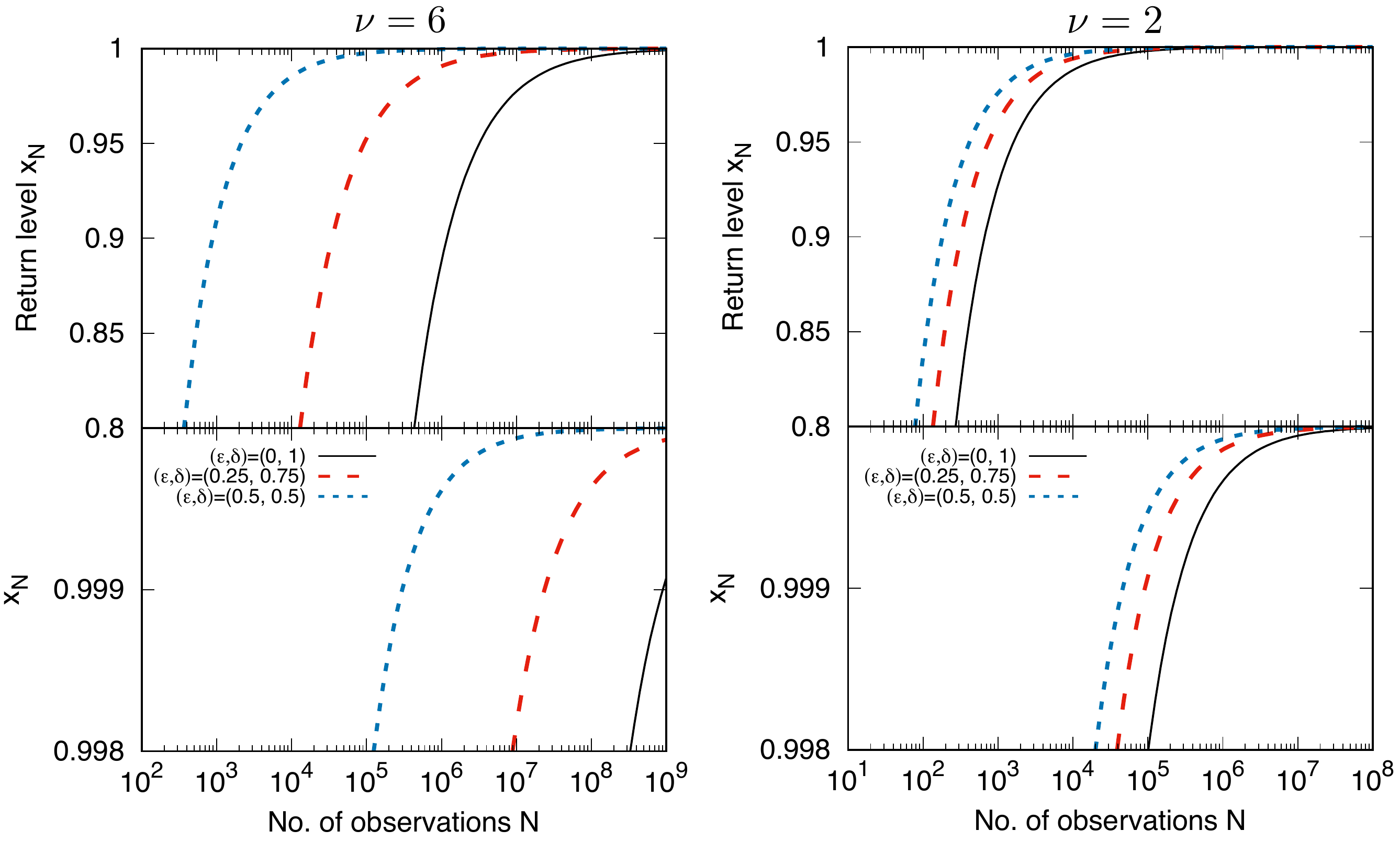}
\caption{The return-level plot, showing the number of PBHs, $N$, amongst which a single one would have formed with spin value exceeding $x_N$, using the parameters $(\nu,\gamma)$, where $\gamma=0.85$ and $\nu=6$ (left) and $\nu=2$ (right), with the two SAS transformations applied.}
\label{figxn2}
\end{center}
\end{figure}

We then repeated the procedure in the previous section to obtain the return levels, $x_N$, for the SAS-transformed spin distributions. The results are shown in Fig. \ref{figxn2}. We see for $\nu=6$ (left panels), the SAS transformations are very effective in boosting the return levels. For instance, the rarity of Thorne-limit violating PBH was reduced from 1 in $\sim3\times10^8$ (for the fiducial model, black curve) to $1$ in $10^5$ when $\eps=\delta=0.5$ (dotted blue curve). For the astrophysical limit $a_s=0.8$, the return level of 1 in $4\times10^5$ is relaxed to 1 in $300$ using the same transformation. By comparison, the improvement in the return levels is not as dramatic for the extreme parameter choice $\nu=2$. 


Our conclusion here is that the return levels can indeed be improved by adjusting the skewness and kurtosis of the underlying spin pdf. A moderate modification of the skewness and kurtosis can boost the chance of a PBH forming with $a_s\gtrsim0.8$ from one in a million, to one in a few hundred. The same transformation can boost the rarity of forming a Thorne-limit violating PBH from one in a hundred million to one in a \ii{hundred thousand}. There is potential here to model more precisely what post-formation interactions do to alter the moments of the primordial spin distribution, which we expect to be significant. For instance, accretion and merger can increase the fraction of black holes at the tail of the spin distribution \cite{dubois, sadowski}, and thus generally driving up the skewness and kurtosis. Fattening of the tail of the spin pdf can also be achieved by more exotic PBH formation mechanisms such as scalar-field fragmentation \cite{cotner} and formation of PBHs through the collapse of cosmic-string cusps \cite{jenkins}.  We leave a more qualitative investigation of such effects for future work.


\section{Conclusions and discussion}

\noindent \ii{Summary:} We have shown how the rarity of extreme-spin PBHs could be quantified using the Peaks-Over-Threshold (POT) approach in Extreme-Value Statistics, adding to a surprisingly small handful of such applications in astrophysics. We have shown how the high-spin tail of the spin pdf could be analysed using the Generalised Pareto Distribution, and from it we calculated the return levels, \ie, the average number of PBHs amongst which a single PBH was formed with spin exceeding a given threshold. Our main results are summarised in Fig. \ref{figxn}, which shows that, using typical parameter values, roughly one in a million PBHs was formed with spin $a_s\gtrsim0.8$, and one in a hundred million formed with spin exceeding the Thorne limit 0.998. These results were derived using the spin pdf at formation time derived by de Luca \etal\ (parametrized by 2 variables, $\gamma$ and $\nu$). We emphasise that our results involving the Thorne limit would be extremely difficult to obtain using the spin distribution on its own without the EVS modelling, because  floating-point errors become overwhelmingly prohibitive at the extreme tail of the distribution.

Furthermore, we found that the return levels of extreme-spin PBHs can be improved by any of the following modifications: a) decreasing $\gamma$, b) decreasing $\nu$ , c) modifying the skewness and kurtosis of the underlying pdf. We demonstrated modification c) by appealing to the \ii{sinh-arcsinh} transformation \cite{jones} which can smoothly alter the skewness and kurtosis of a given pdf. This work, far as we know, is its first application in astrophysics. Moderate changes to both the skewness and kurtosis can change the rarity of forming a PBH with spin exceeding the Thorne limit, from  1 in a hundred million to 1 in a hundred thousand. Fig \ref{figxn2} summarises the SAS  methodology.

\mmm

\noindent\ii{Implications for PBHs as dark matter}: Between formation and the present epoch, there is a potential boost by a factor 
$z\sub{formation} /z\sub{equality}$  depending on BH mass and formation time.  Remarkably, for PBHs in the asteroid mass range, where a possible dark matter window remains open \cite{carr}, the boost in current abundance between formation epoch and the end of the radiation-dominated epoch is comparable to our computed extremal-spin return levels.
The current limit on evaporating PBHs from Voyager-1 $e^{\pm}$
\cite{boudaud} and INTEGRAL data \cite{laha} is that PBHs of mass  $\ltsim  10^{17}\rm g$ 
cannot contribute significantly to the dark matter. However, our estimates imply that a significant number of lower mass PBHs may be stabilised by near-extremal spin and thereby contribute to the dark matter (for example, a $10^{16}\rm g$ PBH formed at $\sim 10^{-23}$s). The surviving extremal PBHs from this epoch  could be as rare as one in $10^{17}$ at formation compared to their evaporating counterparts and yet be  a significant dark matter contributor today. If indeed asteroid mass PBHs contribute to the dark matter, it is likely that some of them,  at somewhat lower masses, could be long-lived near-extremal  PBHs.

\mmm

\noindent\ii{Further astrophysical applications of EVS}: Beyond the application to PBH spin discussed in this work, the POT formalism is applicable to other contexts wherever the rarity of extreme objects is to be quantified, \eg  extreme-mass clusters or extreme-radius cosmic voids. An alternative  extreme-value technique (the `exact' formulation of extreme-value statistics) was applied to these problems in \cite{ mevoid1, mevoid2,chongchitnan}, but it would be interesting to see what the POT formalism could add to previous findings. The SAS transformation, in the case of massive clusters and voids, would also translate directly to primordial non-Gaussianities which affect the skewness and kurtosis of the pdf of primordial overdensities \cite{me1}. Other extensions include the analysis of POT statistics for different black hole spin or charge distributions, and modelling the change in the spin distribution due to evaporation, merger and late-time accretion as redshift-dependent SAS transformations.

We have only investigated the theoretical predictions of PBH spin. However, measuring black hole spin accurately is a monumentally delicate experimental task \cite{reynolds}. Much more will be learnt about black hole spin when the results from the Event Horizon Telescope are further analysed \footnote{\url{https://eventhorizontelescope.org}}. This would give us further clues for the existence and nature of rotating PBHs. Notwithstanding these demanding observational tasks, we believe our work has shed some light on the question  \ii{``how rare are extreme-spin PBH?''}  using Extreme-Value Statistics,  a tool that  deserves to be more widely adopted in astrophysics.


\bibliographystyle{apsrev4-1}
\bibliography{Spin-paper}

\appendix

\section{The probability density function of PBH spin}\lab{appA}

We summarise the analytic form of the pdf of PBH spin at formation time obtained by de Luca \etal\ \cite{deluca}. The spin of a PBH can be quantified by either the Kerr parameter,  $a_s$,  or the spin variable $s_e$. They are related by 
\ba a_s \simeq {0.675 \Omega\sub{DM}\over \pi \nu}s_e,\ea
(although see \cite{harada2} for a different viewpoint on the factor $\Omega\sub{DM}$). The spin pdf is given as a function of $s_e$ by
\ba P(s_e) = {N_1(s_e, \nu, \gamma) \over N_2(\nu,\gamma)},\ea
where 
\ba
N_1(s_e, \nu, \gamma) &= \frac{4\tilde{C}\,s_{\rm e}}{\nu^5}\int_{0}^{\infty} \D \lambda_1 \int_{0}^{\lambda_1} \D \lambda_2 \int_{0}^{\lambda_2} \D \lambda_3 \int_{\alpha_1}^{\alpha_2} \D \beta \frac{e^{-Q_5}\tilde{F}(\lambda_i)\,\Lambda \,T(s_{e}, \nu, \gamma)}{\sqrt{|(\alpha_1^2-\beta^2)(\alpha_2^2-\beta^2)(\alpha_3^2-\beta^2)|}},\\
N_2(\nu, \gamma)  &= \frac{1}{(2\pi)^2}e^{-\frac{\nu^2}{2}} \int_{0}^{\infty}\D x \,\tilde{f}(x) \sqrt{\frac{\Gamma}{2\pi}}e^{-\frac{\Gamma}{2}(x-\gamma\nu)^2}.
\ea
The various components in $N_1$ and $N_2$ are given below.
\bas
\tilde{C} &= \frac{3^{11}5^{11/2}\gamma^5\Gamma^{3/2}}{2^{13}\pi^{13/2}}, \qquad \Gamma = {1\over 1-\gamma^2},\qquad  2Q_5 = \nu^2 + \Gamma(x-x_*)^2 + 15y^2 + 5z^2,\qquad  \Lambda = \lambda_1 \lambda_2 \lambda_3, \\
x&=\lambda_1+\lambda_2+\lambda_3, \qquad y = \frac{1}{2}(\lambda_1-\lambda_3), \qquad z = \frac{1}{2}(\lambda_1-2\lambda_2+\lambda_3),\\
\tilde{F}(\lambda_i) &= \frac{27}{2}\lambda_1 \lambda_2 \lambda_3 (\lambda_1-\lambda_2)(\lambda_2-\lambda_3)(\lambda_1-\lambda_3), \\
\alpha_1 &= \frac{1}{\lambda_3}-\frac{1}{\lambda_2}, \ \ \alpha_2 = \frac{1}{\lambda_3}-\frac{1}{\lambda_1}, \ \ \alpha_3 = \frac{1}{\lambda_2}-\frac{1}{\lambda_1},\\
T(s_e, \nu, \gamma) &= \Theta(\alpha_3^2-\beta^2)
e^{\frac{-15\Gamma w_3^2}{2}}D(X) + 
\Theta(\beta^2-\alpha_3^2)
\frac{\sqrt{\pi}}{2}
e^{\frac{-15\Gamma w_{\beta}^2}{2}}{\rm erf}(X),\\
D(X)&= e^{-X^2}\int_0^X e^{-y^2}\D y , \quad X= \sqrt{\frac{15}{2}\Gamma |w_{\beta}^2-w_3^2|}, \quad w_3=\frac{\sqrt{\Lambda}s_{\rm e}}{K \nu^{5/2}\alpha_3}, \ \ w_{\beta}=\frac{\sqrt{\Lambda}s_{\rm e}}{K \nu^{5/2}\beta}, \ \ K = \frac{2^{9/2}\pi}{5 \times 3^{7/2}\gamma^{5/2}},\\
\tilde{f}(x)&= \frac{(x^3-3x)}{2}\bigg[{\rm erf}\left(x\sqrt{\frac{5}{2}}\right) + {\rm erf}\left(\frac{x}{2}\sqrt{\frac{5}{2}}\right)\bigg] + \sqrt{\frac{2}{5\pi}}\bigg[\left(\frac{31x^2}{4} + \frac{8}{5}\right)e^{-\frac{5x^2}{8}}+ \left(\frac{x^2}{2}-\frac{8}{5}\right)e^{-\frac{5x^2}{2}}\bigg].
\eas
Here $\Theta(x)$ is the Heaviside step function and $\text{erf}(x)$ is the error function.

\end{document}